\documentclass[runningheads,fleqn]{svmult}

\usepackage{makeidx}   
\usepackage{graphicx}  

\usepackage{subeqnar}  
\usepackage{multicol}  

\usepackage{physmult}  
\makeindex             

\begin{document}

\title*{Precision Optical Measurements and Fundamental Physical Constants}

\toctitle{Precision Optical Measurements and Fundamental Constants}

\titlerunning{Precision Measurements and Fundamental Constants}

\author{Savely G. Karshenboim\thanks{E-mail: sek@mpq.mpg.de}}

\authorrunning{Savely G. Karshenboim}

\institute{D. I. Mendeleev Institute for Metrology, St. Petersburg, Russia\\
Max-Planck-Institut f\"ur Quantenoptik, Garching, Germany
}

\maketitle

\begin{abstract}
A brief overview is given on precision determinations of values of
the fundamental physical constants and the search for their
variation with time by means of precision spectroscopy in the
optical domain.
\end{abstract}

\section{Introduction}
The electron and proton are involved in various phenomena of
different fields of physics. As a result, fundamental physical
constants related to their properties (such as electron/proton
charge ($e$), the electron and proton masses ($m_e$ and $m_p$),
the Rydberg constant ($Ry$), the fine structure constant
($\alpha$) etc), can be traced in various basic equations of
physics of atoms, molecules, solid state, nuclei and particles
etc. Until recently most accurate measurements came from radio
frequency experiments only, which supplied us with precise values
of most of the fundamental constants and accurate tests of the
quantum theory of simple atoms (bound state QED). The optical
measurements delivered to us a value for the Rydberg constant, but
were not accurate enough to provide any competitive test of the
QED.

During the last decade the status of the optical measurements
changed dramatically because of
\begin{itemize}
\item advances in atomic spectroscopy, for example, Doppler-free two-photon
spectroscopy of atomic hydrogen, with significantly increased
resolution;
\item advances in the technology for measuring optical frequencies.
New frequency chains routinely deliver the high accuracy of the
microwave cesium radiation (related to the definition of the {\em
second\/}) to the optical domain.
\end{itemize}

Two-photon methods were developed for hydrogen spectroscopy at
Stanford, Oxford, LKB (Laboratoire Kastler Brossel), Yale and MPQ
during two last decades (see a review in \cite{sgk60bj}). 
Improved accuracy of the Rydberg constant
by few orders of magnitude is supplying us with a precision test
of the theory of the Lamb shift in hydrogen and deuterium atoms
(see~\cite{sgk602pho} and references therein). The accuracy of
these tests is even higher than from traditional microwave
measurements. We discuss that in detail in Sect.~2.

The progress with the spectroscopy of the $1s-2s$
transition~\cite{sgk601s2s} and with the development of a new type
of frequency chain~\cite{sgk60chain} was so large, that the
transition offered also an opportunity to search for a variation
of the fundamental constants with time. The appearance of the new
frequency chain developed at MPQ~\cite{sgk60chain} and
successfully applied at several laboratories (MPQ,
JILA~\cite{sgk60JILA}, NIST, PTB) has greatly changed the
situation with the optical measurements (see a review in 
\cite{sgk60uf}). This is important for
metrology and the design of new frequency standards, for the
search of variations of the fundamental constants and for numerous
other applications. We consider the application to the search for
such variations in Sect.~3.

\section{Rydberg Constant and the Lamb Shift in the Hydrogen Atom}
\begin{figure} 
\begin{minipage}[b]{0.40\textwidth}
{\includegraphics[width=\textwidth]{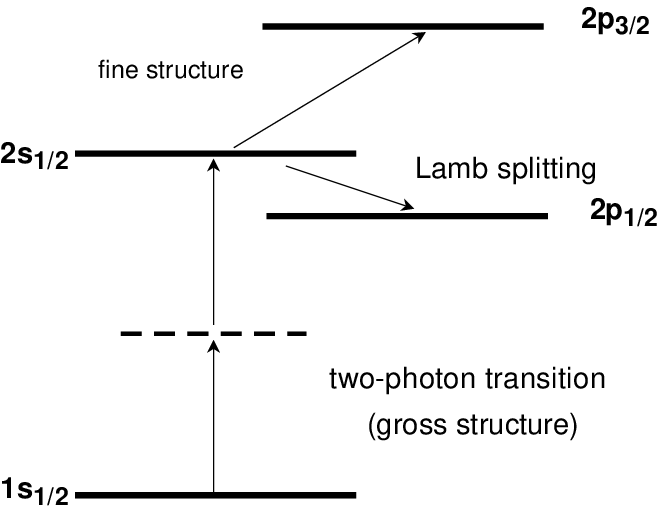}}
\end{minipage}%
\hskip 0.08\textwidth
\begin{minipage}[b]{0.50\textwidth}
{\includegraphics[width=\textwidth]{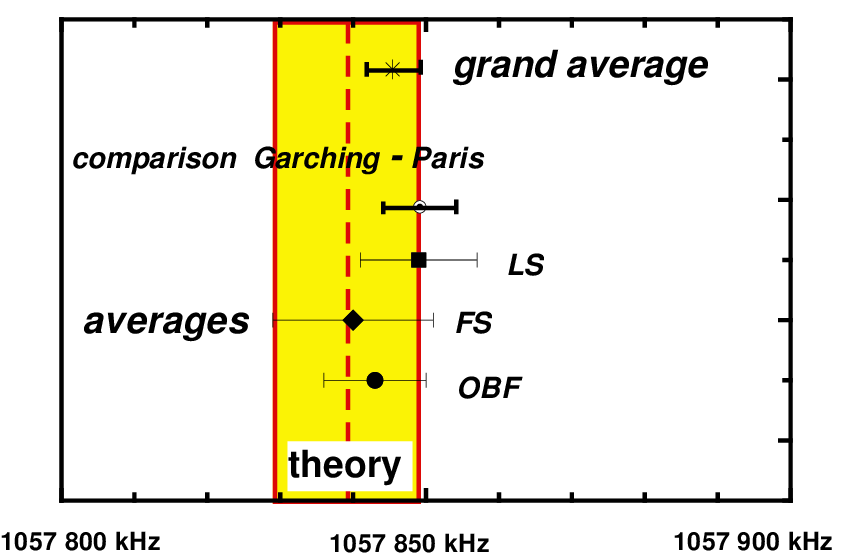}}
\end{minipage}
\vspace{10pt}
\begin{minipage}[t]{0.40\textwidth}
\caption{Example of different transitions in hydrogen atom.
Transitions within the fine structure and Lamb splitting are in
the microwave range, while the $1s-2s$ two-photon transition lies
in ultraviolet domain. The Lamb splitting is the difference of
Lamb shift of $2s_{1/2}$ and $2p_{1/2}$ levels.}

\label{sgk60fi}

\end{minipage}%
\hskip 0.08\textwidth%
\begin{minipage}[t]{0.50\textwidth}
\caption{Present status of the Lamb splitting in hydrogen
($2s_{1/2}-2p_{1/2}$). The figure contains values derived from
various experiments. {\em LS\/} stands for the Lamb splitting
measurements (see Fig.~1), {\em FS\/} is for the fine structure,
and {\em OBF\/} stands for optical beat frequency (simultaneous
measurement of two optical transitions). The theoretical
estimation is taken from~\protect\cite{sgk60Rp}}
\label{sgk60fi2}
\end{minipage}
\end{figure}

The discovery of the Lamb shift in the hydrogen atom was a
starting point of the most advanced quantum theory --- Quantum
electrodynamics (QED). This theory predicts a number of quantities
with a great accuracy. There are only a few examples where an
accurate theory also allows precise measurements. Those
are~\cite{sgk60icap,sgk60h2} the anomalous magnetic moment of the
electron and muon and some transition frequencies in simple atoms
(see e.g. Fig.~\ref{sgk60fi}). Most success was obtained with the
study of the hydrogen atom. However, the progress of traditional
microwave measurements has been quite slow and reached the 10~ppm
level of accuracy only for the hydrogen Lamb shift
(Fig.~\ref{sgk60fi2}).

It turned out that the two-photon Doppler-free spectroscopy of
gross structure transitions (such as $1s-2s$, $1s-3s$, $2s-4s$
etc) allowed access to narrower levels and could deliver very
accurate values sensitive to QED effects. But to interpret those
values in terms of the Lamb shift, two problems had to be solved:
\begin{itemize}
\item the Rydberg constant determines a dominant part of any
optical transition and has to be known itself;
\item a number of levels are involved and it is necessary to be
able to find relationships between the Lamb shifts ($E_L$) of
different levels. Otherwise the experimental data would be of no
use because the number of unknown quantities exceeds the number of
measured transitions.
\end{itemize}

The former problem has been solved by comparison of two
transitions determining both: the Rydberg and the QED
contributions. Presently the two best results to combine are the
$1s-2s$ frequency in hydrogen and
deuterium~\cite{sgk601s2s,sgk60Iso} and the $2s-8s/d$ transition
in the same atoms~\cite{sgk60LKB}. The latter problem has been
solved with a help of a specific difference~\cite{sgk60del}
\begin{equation}
\Delta(n) = E_L(1s)-n^3E_L(ns)\;,
\end{equation}
which can be calculated more accurately than the Lamb shift of the
individual levels.

A successful deduction of the Lamb shift (Fig.~\ref{sgk60fi2}) in
the hydrogen atom provides us with a precision test of bound state
QED and offers an opportunity to learn more about the proton size.
Bound state QED is quite different from QED for free particles.
The bound state problem is complicated itself even in the case of
the classical mechanics. The hydrogen atom is the simplest atomic
system; however, a theoretical result for the energy levels is
expressed as a complicated function (often a perturbative
expansion) of a number of small parameters~\cite{sgk60icap} (see
review~\cite{sgk60egs} for a collection of theoretical
contributions):
\begin{itemize}
\item $\alpha$, which counts the QED loops;
\item the Coulomb strength $Z\alpha$;
\item electron-to-proton mass ratio;
\item ratio of the proton radius to the Bohr radius.
\end{itemize}
Indeed, in hydrogen $Z=1$, the origin of the correction is very
important and the behaviour of expansions in $\alpha$ and
$Z\alpha$ differs from each other. In particular, the latter
involves large logarithms ($\ln(1/Z\alpha)\sim
5$)~\cite{sgk60icap,sgk60jetp} and big coefficients. It is not
possible to do any exact calculations and one must at least use
expansions in some parameters. In such a case the hardest
theoretical problem is not to make the calculation, but rather to
estimate the uncalculated terms related to higher-order
corrections of the expansion.

The other problem is due to the proton size~\cite{sgk60Rp} which
lead to a dominant uncertainty of the theory. The finite size of
the proton leads to a simple expression, but to obtain its
numerical value one needs to determine the value for the proton
charge radius. Unfortunately no such measurements are available at
the moment. It happens now that the most accurate value can be
obtained from a comparison of the experimental value of the Lamb
shift derived from the optical measurements of the hydrogen and
QED theory~\cite{sgk60Rp}.

\begin{figure}[ht] 
\centerline{\includegraphics[width=0.6\textwidth]{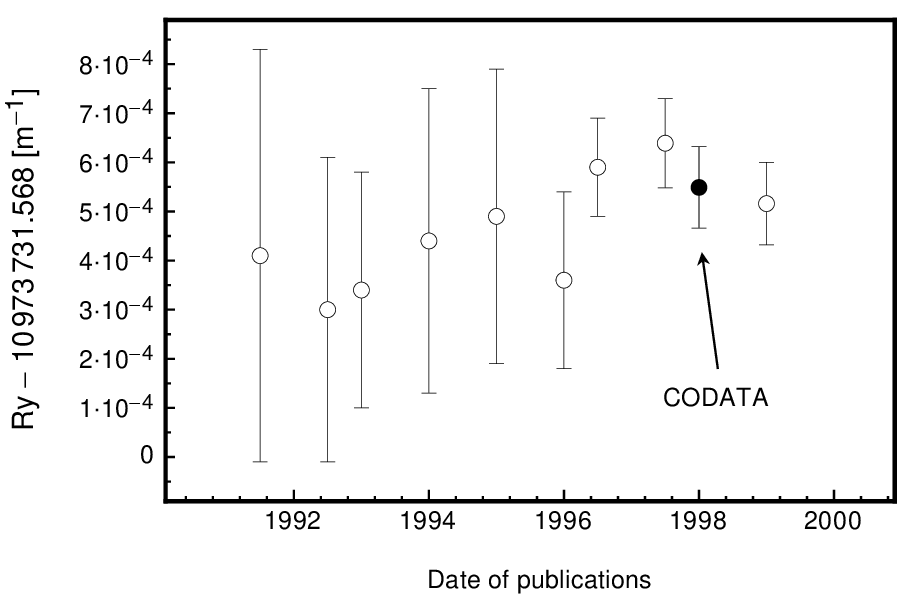}}
\vspace{10pt} \caption{Recent progress in the determination of the
Rydberg constant (mainly at MPQ and LKB)} \label{sgk60fi3}
\end{figure}

The other quantity deduced from the optical measurements on
hydrogen and deuterium is the Rydberg constant. The recent
progress is clearly seen from the recommended CODATA values of
1986 and 1998~\cite{sgk60cod}:
\[
{\rm Ry}_{86} = 10\,973\,731.534(13)\;{\rm m}^{-1}~{\rm and}~{\rm Ry}_{98} = 10\,973\,731.568\;549(83)\;{\rm m}^{-1}\,.
\]
The former value was derived from one-photon transitions (Balmer
series) and was slightly improved later (by a factor of 4.5), but
all further progress that led to the 1998's value (30 times
improvement) was a result of the study of two-photon transitions
in hydrogen and deuterium. Fig.~\ref{sgk60fi3} shows a comparison
of several recently published values for the Rydberg constant.

Some other fundamental constants can also be determined by means
of laser spectroscopy. To complete the overview let us mention
determination of
\begin{itemize}
\item the muon-to-electron mass ratio from three-photon ionization
of the ground state of muonium at a resonance point of the
two-photon excitation of the $2s$ state~\cite{sgk60mu};
\item the fine structure constant $\alpha$ derived from the
helium fine structure;
\item the fine structure constant deduced from Raman recoil spectroscopy
of the cesium $D_1$ line~\cite{sgk60chu}. For this evaluation a
precise value of the absolute frequency of the $D_1$ line is also
necessary.
\end{itemize}

\section{Optical Measurements and Variation of the Fundamental Constants with Time}
After the success with understanding of the electromagnetic, weak
and, in part, strong interaction we arrived at some sort of
threshold of new physics. Naturally, it is not clear on which
front one is most likely to discover new physical phenomena. The
present-day attempts to discover new physics are rather a kind of
a random search for a hidden treasure. Unfortunately, lacking of
theoretical predictions there seems to be no better method up to
now. One of few directions of such a search is related to the
variation of the fundamental constants with time. There is no
common model for such a phenomena, but the accepted picture of the
evolution of our universe strongly implies for that. `Variation'
means a slow drift or oscillation of parameters of the interaction
and particle properties (like their masses). We commonly believe
that during a small fraction of the very first second our universe
came through a number of phase transitions and the interactions
and particles as we know them now did not exist before those
transitions. So, philosophically speaking, we have to acknowledge
the variation of constants as some kind of trace of the early
great changes. Physically speaking, we understand that the
critical question for a detection of the variation is the rate at
which it occurs, because variation rates of $10^{-10}$~yr$^{-1}$
and $10^{-20}$~yr$^{-1}$ are not the same. Variation of, e. g. the
fine structure constant, at the former limit would have already
been detected a few decades ago, while the latter rate will rather
be a challenge for physicists in a few decades. Present-day
searches are related to a level of $10^{-13}-10^{-15}$~yr$^{-1}$,
which corresponds to a fractional shift of the constants smaller
than than $10^{-3}-10^{-5}$ for the life-time of the universe.

There are a number of possibilities for such a search, but the
optical measurements play a specific role, because of their clear
interpretation. However, the fact of variation is more important
than the accurate interpretation. In the case of a negative result
no limitation for a variation can be assigned without a reasonable
interpretation. From the theoretical point of view we have to
expect simultaneous variations of all coupling constants, masses
and magnetic moments. If one wishes to find a solid
interpretation, some reduction of all variable quantities to a
very few is important. However, nuclear properties are the result
of the strong interactions and involve effects, that cannot be
calculated. Only optical transitions are completely free from this
problem.

\begin{table}
\caption{Comparison of different kinds of searches for a variation
of fundamental constants. Here $T$ is a half-period of
oscillations and $\Delta t$ is a time separation for a comparison.
The {\em Laboratory search} contains all laboratory measurements,
while {\em Optical transitions} are for comparison of only optical
transitions. The limitation expected in {\em 1-2 years} are also
presented.} \label{sgk60T2}
\begin{center}
\begin{tabular}{c@{\hskip 2em}c@{\hskip 2em}c@{\hskip 2em}c@{\hskip 2em}c}
\hline
 & Geochemical & Astrophysical & Laboratory & Optical \\
 & study & observation & search & transitions \\
\hline
\hline
Oscillation     &  $\Delta t > T$ & $\Delta t > T$ & phase - ? & phase - ?\\
Space           & $-$ & important & $-$ & $-$ \\
Statistics      & $-$ & essential & & $-$ \\
Strong  Int.    & sensitive & not sensitive & sensitive & not sensitive \\
\hline
$\alpha$        & $+$ & $+$ & $+$ & $+$   \\
$m_e/m_p$       & $-$ & $+$ & $+$ & $-$ \\
$g_p$           & $-$ & $+$ & $+$ & $-$ \\
$g_n$           & $-$ & $-$ & $+$ & $-$ \\
\hline
Limitations & $10^{-17}$ yr$^{-1}$ & $10^{-15}$ yr$^{-1}$ & $10^{-15}$ yr$^{-1}$ & $10^{-13}$ yr$^{-1}$\\
In 1-2 years    &  &  & & $10^{-14}$ yr$^{-1}$\\
\hline
\end{tabular}
\end{center}
\end{table}

We summarize a comparison of different searches in
Table~\ref{sgk60T2}~\cite{sgk60var}. Let us discuss briefly the
specific features of the different methods.
\begin{itemize}
\item To detect a variation one needs to compare some quantity $A(t)$
measured at time $t$ and $t+\Delta t$. However, the obvious
estimation of the variation rate $\Delta A/\Delta t$ is valid only
in the case of a slow drift. In the case of oscillations, and such
oscillations were suggested because of some astrophysical reasons,
the estimate must be rather $\Delta A/T$, where $T$ is the
half-period of the oscillation ($\sim 10^8$ yr). That makes
astrophysical ($\delta t\sim 10^{10}$ yr) and geochemical ($\delta
t\sim 2\cdot10^9$ yr) estimates much weaker, while in the case of
any laboratory experiments we arrive at the question of the phase
of such an oscillation. One sees the laboratory search and a
search over time (and space) are two different kinds of
experiments and serve different purposes.
\item Astrophysical data owe two features different from others. First,
we can observe the object separated from us both in space and time
as there might be a correlation between time- and space-
variations. Secondly, the astrophysical data are hard to interpret
on an event-by-event base. They used to be treated statistically
and it is necessary to study the correlations of the data.
\item Another important question for interpretations is involvement of
the strong interactions. In the case of geochemical data that is
the main effect and there is no reasonable interpretation of such
data at all. In the case of laboratory measurements, the strong
interaction is important because of the nuclear magnetic moments
(see below).
\item We can look only for variations of dimensionless quantities such as
a ratio of two frequencies and we need to reduce them to a
variation of few dimensionless constants, which are the fine
structure constant $\alpha$, electron-to-proton mass ratio,
$g$-factors of proton and neutron. That is possible because of two
reasons, which are
\begin{itemize}
\item known non-relativistic dependence of any transition frequency
on the fundamental constants;
\item the Schmidt model predicts nuclear magnetic moments for odd
$Z$;
\end{itemize}
\item Non-relativistic theory in particular predicts (see~\cite{sgk60var}
for details), that
\begin{itemize}
\item any gross structure transition frequency depends only on the
Rydberg constant ($\sim Ry$);
\item any fine structure transition frequency is proportional to
$\alpha^2 Ry$;
\item any hyperfine structure transition is proportional to $\alpha^2
(\mu/\mu_B) Ry$, where $\mu$ is the nuclear magnetic moment and
$\mu_B$ is the Bohr magneton.
\end{itemize}
\item The Schmidt model predicts the magnetic moment of nuclei with
odd mass number $A$ as a result of the spin and orbit contribution
of a single nucleon (the remaining protons and neutrons are
coupled in pairs and do not contribute). Indeed that is a very
rough model, but there is no other way to reduce all magnetic
moments to few quantities ($m_e/m_p$, $g_p$, $g_n$). The problem
of interpretation is now related to the inaccuracy of the Schmidt
model because of the strong interaction which is not under
control.
\end{itemize}

The present-day level of limitations is different for various
methods and the optical experiments do not look as a good choice
in general and in comparison with other laboratory experiments.
However, the discussion above explains in part that there is no
clear interpretation of the geochemical data and there are some
doubts in the astrophysical data. There is a single laboratory
result in Table~\ref{sgk60T2} that is better than $10^{-13}$
yr$^{-1}$: a comparison of the hyperfine splitting in Rb and
Cs~\cite{sgk60Rb}. This experiment could be a search for a
variation of $g_p$. Such an interpretation is valid only if the
Schmidt model can be applied, but unfortunately there are
essential deviations from the Schmidt model due to the strong
interaction (mainly in Cs). The actual interpretation remains
unclear.

We briefly discuss optical laboratory experiments below. The
current limitations are quite weak ($\sim 10^{-13}\;{yr}^{-1}$)
because most of the results were obtained recently on a level of a
fractional uncertainty of $10^{-14}$ (see Table~\ref{sgk60T1}).
These limitations were obtained either after a short-term
monitoring or after a comparison with a previous less accurate
result. Future limitations based on reproduction of the recent
experiments in 2000-2001 must easily deliver limitations better by
an order of magnitude.

\begin{table}
\caption{Some results of recent precision optical measurements (H,
$1s-2s$, $^{40}$Ca, $^3P_1-{^1}S_0$, $^{115}{\rm In}^+$,
$5s^2{^1}S_0-5s5p{^3}P_0$, $^{171}{\rm Yb}^+$,
$6s^2{^2}S_{1/2}-5d^2{^2}D_{3/2}$, $^{199}{\rm Hg}^+$,
$^2S_{1/2}-{^2}D_{5/2}$), their fractional uncertainty $\delta$
and their sensitivity to a variation of the fine structure
constant $\kappa =
\partial\ln{F_{rel}(\alpha)}/\partial\ln{\alpha}$} \label{sgk60T1}
\begin{center}
\begin{tabular}{c@{\hskip 2em}r@{\hskip 2em}c@{\hskip 2em}c@{\hskip 2em}c}
\hline
Atom & Frequency~~~~~~~~~~~~ & Place and date & $\delta$ & $\kappa$\\
 & [Hz]~~~~~~~~~~~~~~~~ & of measurement& [$10^{-14}$]& \protect\cite{sgk60Fla} \\
\hline
H  & 2\,466\,061\,413\,187\,103~(46) & MPQ, 1999, \protect{\cite{sgk601s2s}} & 1.8& 0.00 \\
Ca  & 455\,986\,240\,494\,158~(26) & NIST, 2000, \protect{\cite{sgk60NIST}} & 5.7& 0.03 \\
${\rm In}^+$ & 1\,267\,402\,452\,899\,920(230) & MPQ, 1999, \protect{\cite{sgk60In}} & 18 & 0.21 \\
${\rm Yb}^+$ & 688\,358\,979\,309\,312~~(6) & PTB, 2001, \protect{\cite{sgk60PTB}} & 0.9& 1.03 \\
${\rm Hg}^+$ & 1\,064\,721\,609\,899\,143~(10) & NIST, 2000, \protect{\cite{sgk60NIST}} & 0.9& -3.18 \\
\hline
\end{tabular}
\end{center}
\end{table}

Since all the transitions in Table~\ref{sgk60T1} are related to
the gross structure one could wonder how to obtain any information
about a variation of the fine structure constant if all of them
are proportional (in the non-relativistic approximation) to the
Rydberg constant. The signal is due to the relativistic
corrections. Their importance (for the hyperfine structure) was
first pointed out in~\cite{sgk60Pre} and they were later
calculated for a bunch of the optical transitions
in~\cite{sgk60Fla}. The transition frequency is now equal to
$c\,Ry\,F_{\rm rel}(\alpha)$ and a non-trivial relativistic factor
$F_{\rm rel}(\alpha)$ is a key point to look for a variation of
the fine structure constant by optical means. The sensitivity
$\kappa =
\partial\ln{F_{rel}(\alpha)}/\partial\ln{\alpha}$ to a variation
of $\alpha$ is given in Table~\ref{sgk60T1} accordingly
to~\cite{sgk60Fla}.

A comparison of two optical frequencies can be performed directly
(like e.g. an Hg-Ca comparison at NIST~\cite{sgk60NIST}) or
indirectly (via a comparison of both frequency to cesium
standard).

A direct comparison of two distinct optical frequencies is now
possible using the newly developed femtosecond frequency chains.
This approach has the advantage of a high short-term stability as
compared with cesium standard. This allows for a simple {\em time
structure} of the experiment~\cite{sgk60var} with a few direct
comparisons. A comparison with cesium often involves some
secondary standard with a high short-term stability. Such a clock
being an artefact that is not related to any transition should
involve an unknown drift with time as the constants are drifting.
Their short time stability cannot be actually proved. A common
reason for a statement on the good short-time stability of such a
standard (like a hydrogen maser for example) is that the
scattering of their frequencies with time is small. But there is
no idea about a possible common mode rejection. The stronger
reason to believe in a good short term stability is a comparison
between different clocks. But still there might be a common mode
rejection as it should be the case for a variation of the
constants. A direct comparison of two optical transitions allows
to avoid this problem and offers a new opportunity to securely
derive a limitation for a variation of the fine structure constant
or maybe even to detect such a variation.

\section{Summary}
Nowadays, optical spectroscopy of atoms provides an essential
input for the determination of the fundamental physical constants,
including the most accurately known fundamental constant, the
Rydberg constant, which plays an important role in atomic physics.
The so-called atomic unit of frequency and energy are related to
this constant being $c\cdot Ry$ and $h\cdot c\cdot Ry$
respectively. A definition of the {\em second\/}, attractive from
a general point of view, could be based on a fixed value of the
Rydberg constant. This would be practically acceptable after the
following steps are achieved:
\begin{itemize}
\item the completion of a frequency chain that connects the
optical with microwave domain and, at the same time allows for a
comparison of any optical transition with the $1s-2s$ frequency on
in hydrogen (done at MPQ~\cite{sgk60chain} and used now also at
NIST, JILA, PTB);
\item accurate measurement of the $1s-2s$ transition (done at
MPQ~\cite{sgk601s2s} with an accuracy compatible with any other
optical transition but not yet competitive with cesium and
rubidium fountain clocks);
\item proper QED theory of the Lamb shift (essentially developed
within the last decades but it still needs more progress);
\item determination of the proton charge radius (not known with
sufficient accuracy, but a promising experiment is in
progress~\cite{sgk60PSI}).
\end{itemize}
The hydrogen atom was a candidate for the primary frequency
standard in the 1960ies because of the hyperfine splitting, but
this attempt failed for a number of reasons. Now it is time for a
strong competition of new frequency standards and the hydrogen
atom has its second chance.

Actually there is one more fundamental constant, which is
associated with the optical measurements. That is the speed of
light which is equal to
\begin{equation}\label{sgk-c}
c = 299\,792\,458 \;{\rm m/s}
\end{equation}
because of the definition of the {\em meter\/}. Despite this
constant is fixed by definition there is still an uncertainty
related to a practical realization of the {\em meter\/}. The
problem is that the {\em second} is defined with the help of the
cesium hyperfine transition, while the {\em meter\/} is related to
the optical domain. The accepted recommendations for its
realization is based on optical transitions~\cite{sgk60met}. In
the case of any direct application of (\ref{sgk-c}) one meets
three sources of uncertainty
\begin{itemize}
\item a realization of the {\em second\/} in the microwave domain;
\item a realization of the {\em meter\/} in the optical domain;
\item a bridge between both realizations to the same transition
(i.e. the frequency chain).
\end{itemize}
After recent progress in frequency chain metrology and a long-term
development of the microwave standards, present-day limitations
come mainly from the optical standards, which however are
developing in a very promising way.

The old-fashion frequency chains, which really presented the state
of art in the field just few years ago, appear now as some kind of
dinosaur, which should very soon disappear. Those chains are much
bigger, much more expensive, much more complicated in construction
and in use and at the same time less accurate, than the new
frequency-comb chains. The main disadvantage of the old technology
was a limited possibility in use. The old chains were designed as
a single-problem chain and it was necessary to redesign it with
adding some more items to adjust it to new transitions. The new
chain provides us with a possibility to measure any optical
transitions and some measurements like the cesium $D_1$ line for
example needed for the determination of the fine structure
constant~\cite{sgk60chu} are now a routine problem. The chain
offers new horizons for precision optical spectroscopy and we look
forward for soon reproductions of recent
experiments~\cite{sgk601s2s,sgk60NIST,sgk60PTB} which must deliver
new secure limitations for a possible variation of the fine
structure constant with time on a level of $10^{-14}$ per year.

\section*{Acknowledgement}

I met Ted H\"ansch for the first time at ICAP'94 and became a
frequent visitor to the MPQ. I have been always impressed by the
free and fruitful atmosphere of his lab. Starting as a pure
theorist I am now trying to be between theory and experiment and I
enjoyed lots of discussions with Ted H\"ansch and his
collaborators, which were really stimulating. During the 8 years
around the MPQ, a highlight of our cooperation certainly was the
organization of the {\em Hydrogen Atom, 2\/}
meeting~\cite{sgk60h2}, in which both of us were involved. I was
always feeling his support without which the project could not be
realized. Personally, he did only a few adjustments. But those
were the crucial things that I had missed, and that was one more
chance to see how efficient he is. I do really wish to thank him
for the support and hospitality he extended in Garching and
especially in Florence.

\end{document}